\documentclass[a4paper,11pt]{article}
\pdfoutput=1 
\usepackage{jinstpub} 

\usepackage{subfig}
\usepackage{lineno}
\usepackage{multirow}
\usepackage{blindtext}
\title{\boldmath On a new environment-friendly gas mixture for Resistive Plate Chambers}

\author[a,b,1]{G. Proto,\note{Corresponding author.}}
\author[b]{B. Liberti,}
\author[a,b]{R. Santonico,}
\author[a,b]{G. Aielli,}
\author[a]{P. Camarri,}
\author[b]{R. Cardarelli,}
\author[a,b]{A. Di Ciaccio,}
\author[a]{L. Di Stante,}
\author[c]{A. Paoloni,}
\author[b]{E. Pastori,}
\author[b]{L. Pizzimento,}
\author[b]{A. Rocchi,}

\affiliation[a]{University of Rome Tor Vergata,\\Via della Ricerca Scientifica 1, Rome, Italy}
\affiliation[b]{INFN Rome Tor Vergata,\\Via della Ricerca Scientifica 1, Rome, Italy}
\affiliation[c]{INFN Laboratori Nazionali di Frascati (LNF),\\Via Enrico Fermi, 54, Italy}

\emailAdd{giorgia.proto@roma2.infn.it}
\emailAdd{barbara.liberti@roma2.infn.it}

\abstract{This paper studies the performance of RPCs working with a new family of environment-friendly operating gases, mainly based on Carbon Dioxide and Hydro-Fluoro-Olefins. The tests are carried out on a 2 mm gap RPC and concern the measurement of detection efficiency, avalanche-to-streamer transition probability, prompt and ionic charge delivered. The timing properties of the new gas are also measured.}

\begin{document}
\maketitle
\flushbottom
\newpage
\section{Low Global Warming Potential and Ozone Depletion Potential gases for Resistive Plate Chambers}
\label{sec:intro}
The Resistive Plate Chambers are gaseous detectors characterized by the uniform electric field generated by two parallel electrode plates of high bulk resistivity \cite{a}. They usually work in avalanche mode which is achieved  with a “standard” gas mixture composed of $\rm{C}_{2}$$\rm{H}_{2}$$\rm{F}_{4}$/i-$\rm{C}_{4}$$\rm{H}_{10}$/$\rm{SF}_{6}$ = 94.7/5.0/0.3. 

This gas mixture, which is the result of a relevant R\verb &D  investment \cite{c}\cite{e}\cite{d}, is characterized by a number of properties that make it almost ideal for an excellent RPC performance:
\begin{itemize}
	\item {high gas density ensuring sufficient primary ionization even for gas gaps in the millimeter range size;}
	\item{prompt charge, dominated by the electron drift velocity, slowly increasing with the applied voltage and high enough to easily overcome the front-end electronics threshold\cite{toppi};}
	\item{total charge, dominated by the slow ion drift motion constituting about 90\% of the charge delivered in the gas, low enough to ensure modest working current and good rate capability, as required by high-luminosity collider experiments;}
	\item{comfortable avalanche-streamer separation which allows streamer-free avalanche operation \cite{c};}
	\item{non-flammable and made of industrial components as required for very large size detectors also operating in underground laboratories.}
\end{itemize}
The standard RPC gas is mostly based on a Hydrofluorocarbon, HFC, extensively used in the refrigeration industry. The Hydrofluorocarbons are now considered to be  non-eco-friendly gases for their high Global Warming Potential (GWP) \cite{x}\cite{y}\cite{u}.  In particular, the main RPC gas component, the $\rm{C}_{2}\rm{H}_{2}\rm{F}_{4}$ molecule (Suva 134a) has a GWP of about 1450 and will not be recommended for industrial uses in the next future. In this paper we study new gas mixtures based on Hydro-Fluorine-Olefin molecules, HFOs, which have a GWP around the unity and will replace the HFCs \cite{santpechino}. \\Another component of the RPC gas mixture is the $\rm{SF}_{6}$ molecule, which has a crucial importance as a gaseous insulator for high-voltage power plants, but has the highest GWP $\sim$ 23900 among the industrial gases presently in use. Although the RPC gas contains a very small amount of $\rm{SF}_{6}$, around 0.3\%, finding  a substitute also for this component is desirable anyway.
\\This paper is a systematic study of HFO-based gases \cite{ghent} \cite{Abbrescia:2016xdh}\cite{rpc2020}\cite{pisano}  done with the purpose of replacing the HFC components without losing the excellent properties of the standard gas. It should be stressed that this replacement is somewhat more critical for the RPCs already working in the ATLAS, CMS and ALICE experiments at the CERN Large Hadron Collider (LHC), with respect to new RPCs, to be conceived and installed during phase-II upgrade. In the latter case indeed the new chamber setup, gap size, front-end electronics etc can be optimized to account for the new gas properties. This is not possible for the chambers already installed, which cannot be upgraded because their dis-installation looks particularly difficult. \\HFO gases however, when used in RPCs, exhibit important differences in comparison with the $\rm{C}_{2}\rm{H}_{2}\rm{F}_{4}$ gas mixture used so far. In particular, they require much higher electric fields. It increases the charge per pulse delivered in the gas thus reducing the detector rate capability and might also accelerate the detector ageing.
Therefore in performed tests the HFO is always mixed with $\rm{CO}_{2}$, which becomes the main component of the RPC gas.

 \section{Experimental test of HFO-$\pmb{\rm{CO}_{2}}$ based gas mixtures }
 \label{sec:experimental_test}
\begin{figure}[h]
 	\centering
 	\includegraphics[width=0.4\textwidth]{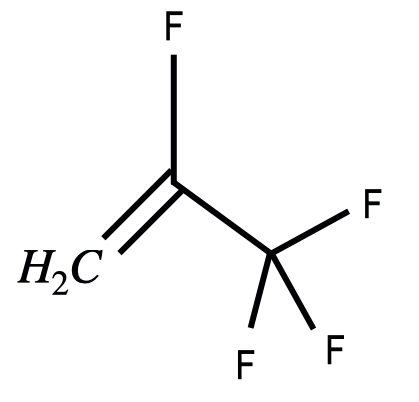}
 	\caption{1,3,3,3-Tetrafluoropropene molecule.}
 \label{fig:hfof_mol}
 \end{figure}

We summarize in this section some results already published on  $\rm{CO}_{2}$/HFO mixtures \cite{SIF2020}. 
 The HFO1234ze molecule, shown in Figure \ref{fig:hfof_mol} is a type of Tetrafluoropropene. Among industrial HFOs it is the most similar to the HFC used in the standard mixture.  Therefore it was the first HFO to be tested\cite{newmixcarda}. It should be stressed that this is the first molecule characterized by a double C=C bond used as sensitive target for gaseous detectors. For semplification we will refer to this molecule as F-HFO in the following. 

Several four-component mixtures of F-HFO/$\rm{CO}_{2}$/i-$\rm{C}_{4}\rm{H}_{10}$/$\rm{SF}_{6}$ have been studied in the range of F-HFO content between 5\% and 25\%.  In the first set of measurements the i-$\rm{C}_{4}\rm{H}_{10}$ and $\rm{SF}_{6}$ concentrations are fixed at 5\% and 1\% respectively and the ratio F-HFO/$\rm{CO}_{2}$ ranges from 5/89 to 25/69. The results, summarized in Table \ref{tab:sif}, show that increasing the F-HFO concentration up to 15\% reduces the delivered charge and improves the streamer discrimination. On the contrary no significant change is observed above 15\% up to 25\% concentration. The optimal choice of F-HFO concentration in this range has to take into account, in addition to the delivered charge, other features like for example the ageing properties.
In the second set of measurements the F-HFO and $\rm{SF}_{6}$ concentrations are fixed to 5\% and 1\% respectively, whereas the i-$\rm{C}_{4}\rm{H}_{10}$/$\rm{CO}_{2}$ ratio is increased up to 15/79 disregarding the flammability limit, with the only purpose of testing the  i-$\rm{C}_{4}\rm{H}_{10}$ properties. The results, summarized in Table \ref{tab:b}, show that the increase of i-butane drastically reduces  both the streamer fraction and total charge,  suggesting that its relative fraction in the mixture should be as high as possible within the flammability limit imposed by safety requirements. Therefore in this paper we specifically test the mixture containing 15\% HFO and 7\% i-$\rm{C}_{4}\rm{H}_{10}$, in the assumption that: the minimum amount of F-HFO should reduce the ageing and the 7\% i-$\rm{C}_{4}\rm{H}_{10}$ is just below the flammability limit. Small readjustment of this limit should not change the following results in a significant way. 

\begin{table}[ht]
  \caption{Knee voltage, streamer fraction, ionic and prompt charges for four-component mixtures  of F-HFO/$\rm{CO}_{2}$/i-$\rm{C}_{4}\rm{H}_{10}$/$\rm{SF}_{6}$ with the i-$\rm{C}_{4}\rm{H}_{10}$ and $\rm{SF}_{6}$ concentrations fixed at 5\% and 1\% respectively and  HFO/$\rm{CO}_{2}$ ratio ranging from 5/89 to 25/69. The streamer fraction, the ionic and the prompt charges are measured at HV = ($\rm{V_{knee}}$+ 200) V. }
 
  \begin{tabular*}{\textwidth}{ccccc}
    \hline
     \% HFO  & $\rm{V_{knee}}$ (kV)  & Streamer fraction & Ionic charge (pC)  & Prompt charge (pC) \\
    
    \hline  
     5 &8.5  & 8.5\% & 50 & 11.5\\
    10 & 9 & 3\% & 38 & 7\\    
   15 & 9.5 &0.6\% &32& 4.3 \\
   20 & 9.9 & 0.8\% &31& 4\\
   25 & 10.4 &0.7\%&34& 5\\
    \hline
         \label{tab:sif}
  \end{tabular*}
 \end{table}
 
 \begin{table}[ht]
  \caption{Knee voltage, streamer fraction, ionic and prompt charges for four-component mixtures of $\rm{HFO}$/$\rm{CO}_{2}$/i-$\rm{C}_{4}\rm{H}_{10}$/$\rm{SF}_{6}$ with the HFO and $\rm{SF}_{6}$ concentrations fixed at 5\% and 1\% respectively and i-$\rm{C}_{4}\rm{H}_{10}$/$\rm{CO}_{2}$ ratio in the range 5/89 - 15/79. The streamer fraction, the ionic and the prompt charges are measured at HV = ($\rm{V_{knee}}$+ 200) V.}
      \label{tab:b}
  \begin{tabular*}{\textwidth}{ccccc}
    \hline
     \% i-$\rm{C}_{4}\rm{H}_{10}$  & $\rm{V_{knee}}$  (kV)  & Streamer fraction & Ionic charge (pC)  & Prompt charge (pC)  \\

    \hline  
     5 &8.5  & 8.5\% & 50&11.5\\
    10 & 8.45 & 3.5\% & 37&7.3\\    
   15 & 8.4 & 1 \%&30 &5.3\\
   
    \hline
  \end{tabular*}
 \end{table}

\begin{figure}[h]
 	\centering
 	\includegraphics[width=0.6\textwidth]{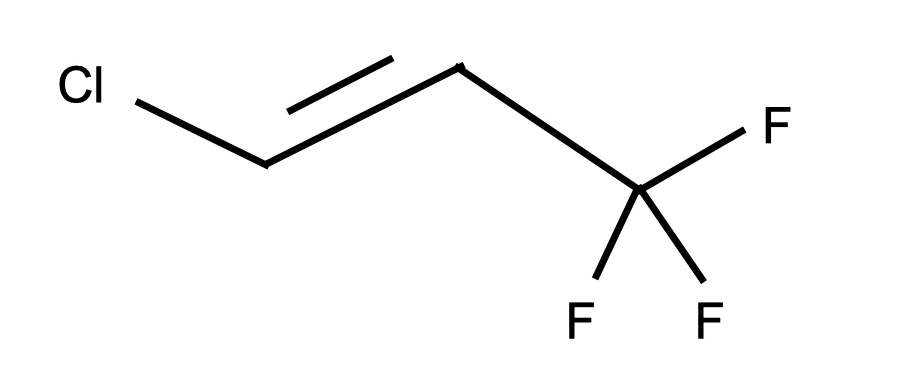}
 	\caption{1-Chlorine-3,3,3-Trifluoropropene molecule.}
 \label{fig:hfoc_mol}
 \end{figure}

Finally in order to obtain a totally eco-friendly gas, a suitable substitute of $\rm{SF}_{6}$ is needed. Previous tests suggest that HFC molecules with two different alogen atoms have a strong quenching effect reducing the avalanche and streamer charge \cite{z}. This type of molecules however show in most cases an elevated Ozone Depleting Power (ODP). A relevant exception is the 1-Chlorine-3,3,3,trifluoropropene (HFO1233zd), shown in Figure \ref{fig:hfoc_mol}, with a GWP $\sim$ 6 and ODP $\sim$ 0 \cite{tabella}. 
We will refer to this molecule as Cl-HFO in the following.\\

To summarize, the search for new eco-friendly gases replacing the standard mixture is carried out following these criteria:

\begin{itemize}
	\item {The Tetrafluoroethane, which is the standard-mixture main component, is replaced by the $\rm{CO}_{2}$/F-HFO mixture. This increases the number of components from 3 to 4;}
	\item{The i-$\rm{C}_{4}\rm{H}_{10}$ is kept at approximately the same proportion as in the standard mixture. Only a very modest increase is applied on the assumption that a large concentration of $\rm{CO}_{2}$ would increase the gas flammability threshold \cite{flam};}
	\item{ The Cl-HFO is studied in details as $\rm{SF}_{6}$ substitute, for suppressing the avalanche-to-streamer transition in the full-efficiency range of operating voltages.}
\end{itemize}

\section{Experimental apparatus}
\begin{figure}[!h]
 	\centering
 	\includegraphics[width=0.3\textwidth]{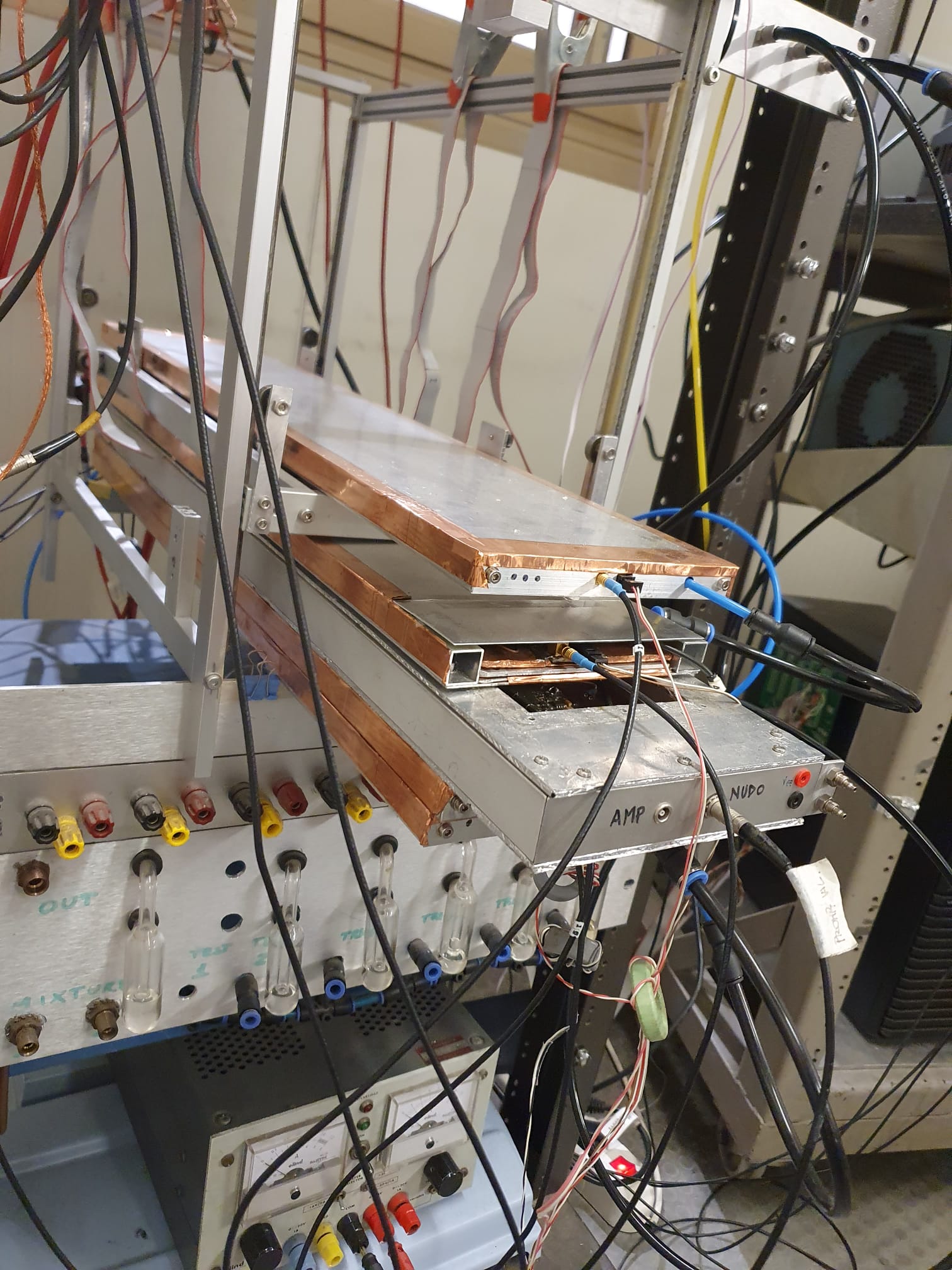}
 	\caption{Picture of the experimental setup.}
 \label{fig:fotosetup}
 \end{figure}

In this section the experimental apparatus, shown in Figure \ref{fig:fotosetup}, is described.

\label{sec:experimental_setup}
\subsection{Test chamber and trigger system}
\label{sec:test_chamber}
\begin{figure}[!h]
\centering
\subfloat[][]{\includegraphics[width=0.95\textwidth]{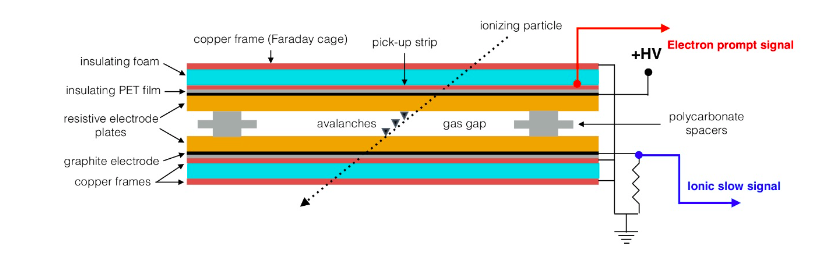} } \\
  \subfloat[][]{\includegraphics[width=0.7\textwidth]{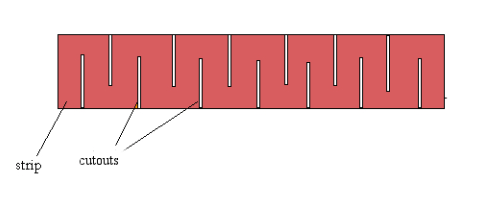} }
 	\caption{Scheme of the test chamber (a) and scheme of the readout strip (b).}
 \label{fig:setup}
 \end{figure}

The test chamber, Figure \ref{fig:setup}.a, is a small-size RPC, $55\times10$ $\rm{cm}^{2}$ area, with 2 mm gas gap and 1.8 mm electrode thick made of phenolic resine based High Pressure Laminate (HPL). Both the gas gap and the electrode thickness reproduce the layout of the RPCs already operating in the ATLAS experiment at the LHC. The graphite electrodes have surface resistivity of 100 $\rm{k\Omega}$ per square. The one connected to ground is used for detecting the ionic signal, by sampling the voltage across a 10 $\rm{k\Omega}$ resistor connecting the  graphite electrode to ground.

The prompt-signal pick-up is made with a single strip, shaped as shown in Figure \ref{fig:setup}.b, of 30 $\Omega$ line impedance covering a pick-up area 35 mm wide, which fully intercepts the solid angle covered by the trigger system. The particular shape reduces the effective width of the transmission line with respect to 35 mm, in order not to lower its characteristic impedance substantially below 30 $\Omega$. \\

The trigger system consists of four RPC detectors selecting cosmic-ray tracks. One of them has a 0.5 mm gas gap and provides a precise time reference ($\rm{\sigma_{time}}$ < 0.5 ns) for the acquired data. The trigger signals are amplified and discriminated. 
Then, a logical unit performs the logic AND of the 4 trigger signals. The trigger rate is 0.5 Hz, which is compatible with the expected rate on the acceptance area (50 × 35 $\rm{cm}^{2}$) and covered solid angle.

 \subsection{DAQ system}
 \label{sec:DAQ}
For each trigger signal the waveforms produced by the test chamber are recorded on a 4-channel oscilloscope with analog bandwidth of 4 GHz and sampling rate of 20 Gs/s. This data acquisition allows storing the maximum amount of information concerning the electrical discharge growing inside the gas, for any subsequent analysis. The prompt signal, generated by the electron drift motion inside the gap, is acquired on both ends of the read-out strip in order to record it with different sensitivity scales of the scope. The strip is terminated on both ends with its own 30 $\Omega$ impedance which requires introducing a 75 $\Omega$ resistor in parallel to the 50 $\Omega$ cable connecting the strip to the scope input. The following signals are recorded for each triggered event:
\begin{itemize}
 
 \item{Prompt signal with the highest sensitivity scale of the scope  in order to detect small signals with the best signal-to-noise ratio, as required for the detection efficiency measurement;}
 \item{Prompt signal with a variable scope scale sensitivity in order to minimize the saturation risk for very large signals generated at operating voltages near to or beyond the streamer appearance. Prompt signals are acquired in a 200 ns  time window, both for high and low sensitivity scale. This wide window allows detecting unwanted events like streamers if any, occurring much after the first short (few ns) avalanche signal;}
 \item{Ionic signal generated by the slow drift motion of the ions inside the gas. This signal is read out on a 10 k$\rm{\Omega}$ resistor connecting the graphite electrode to ground, of 100 $\rm{k\Omega}$ per square surface resistivity. A thin metal wire is embedded in the graphite at the electrode edges for minimizing the signal diffusion time, which would abnormally increase the ionic signal duration. The acquisition window in this case is 100 $\rm{\mu}$s;}
 \item{The prompt signal of the 0.5 mm gap trigger chamber is used to confirm the trigger over a reduced covered area and is assumed as time reference for any timing analysis, due to its higher time resolution.}
\end{itemize}
These four waveforms are recorded for any off-line analysis. For each tested gas mixture a voltage scanning is carried out, from near-to-zero efficiency up to the streamer appearance in steps of 100-200 V, with 300 events acquired at each step. For each operating voltage the acquired data allows to measure: detection efficiency, prompt and ionic charge, streamer
probability and also to detect anomalous afterpulses produced in the 200 ns window. The offline data analysis has been performed using the ROOT framework \cite{root}.

\subsection{Gas system}
\label{sec:gas}
\begin{figure}[!h]
 	\centering
 	\includegraphics[width=0.8\textwidth]{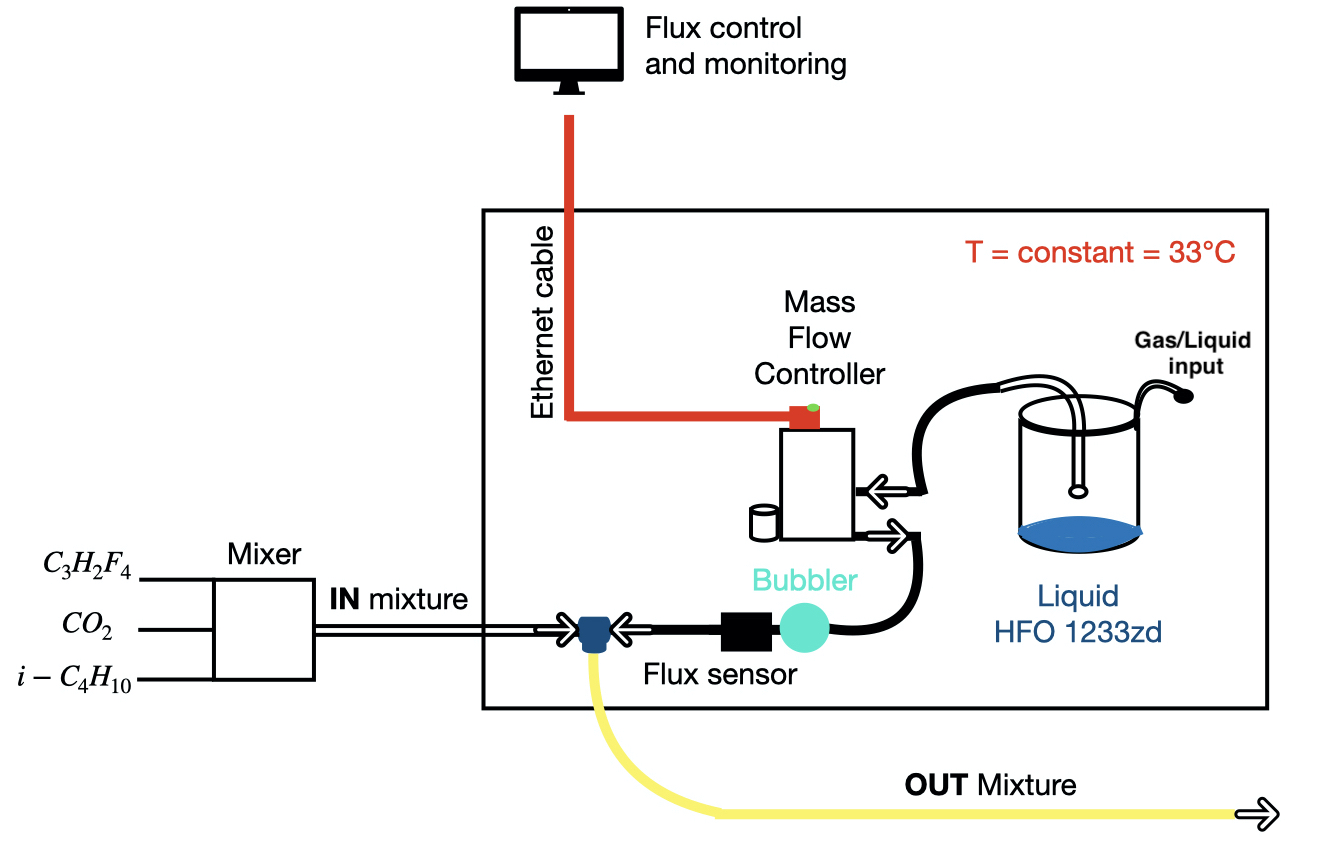}
 	\caption{Gas system scheme.}
 \label{fig:gassystem}
 \end{figure}

The gas mixtures are obtained with a standard gas system, composed by mass flow meters (MFM) and mass flow controllers (MFC). A special attention was dedicated to the flow measurement of the Cl-HFO whose boiling point, T=19 \textdegree C, is very close to the room temperature. In order to avoid any risk of gas liquefation, a small Cl-HFO glass bottle was kept at the controlled temperature of  T=33°C together with the relative MFC. The scheme of the gas system is shown in Figure \ref{fig:gassystem}. 

The gas pressure inside the bottle, about 1,7 atm, is sufficient to keep a constant flow. Outside the MFM the gas flows through a bubbler and an independent flow sensor. The Cl-HFO is mixed with the other three components (F-HFO/$\rm{CO}_{2}$/i-$\rm{C}_{4}\rm{H}_{10}$) inside the environmental chamber. The resulting gas mixture, once outside of the climatic chamber, can be cooled without liquefaction risk. As an example in the mixture with 2\%  of Cl-HFO, liquefaction occurs at temperature much below -10\textdegree C \cite{liq}.

The absolute flux of the new component is evaluated with the flow sensor coupled with the bubbler. 

The total flux inside the test gas volume is of the order of 400 bubbles/min ($\sim$ 60 cc/min).


\subsection{Waveforms analysis and extracted parameters} 
For each triggered event a waveform of 200 ns is acquired for the electron-induced prompt signal. A background window of 30 ns is defined, which is supposed to be empty, together with a pulse window of 40 ns in which the signal is expected and searched. 
 In the background interval the average amplitude of the background level is measured and subtracted to the full waveform to eliminate low-frequency noise. On the other hand, the standard deviation is measured in the same interval in order to set a 5 sigma threshold for the discrimination of the genuine signals. The associated avalanche charge is integrated and normalized with respect to the line impedance Z of the read-out strip, in a 10 ns time interval around the maximum. 
 The total prompt charge, including the afterpulses and streamers is measured by integrating the signal on a 170 ns wide window starting after the background time interval.

The corresponding ionic charge is measured by integrating the ionic signal, detected on a 10 k$\Omega$ resistor connected to the ground, on a  100 $\rm{\mu}$s time window.

\subsection{Analysis criteria and event definition}
\label{sec:par_def}
\begin{figure}[!h]
  \subfloat[][]{\includegraphics[width=0.33\textwidth]{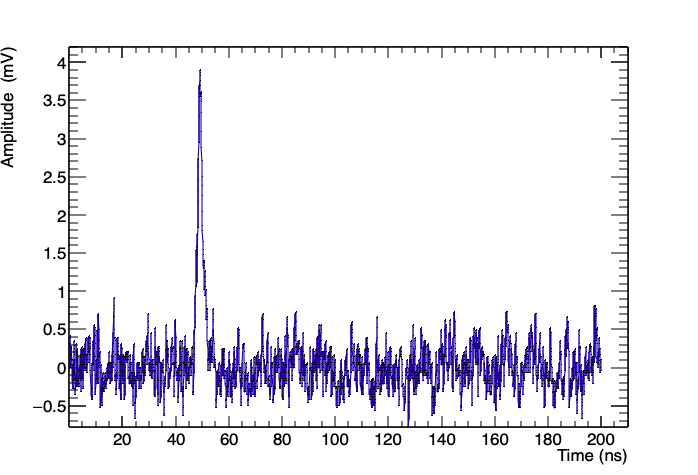} } \quad
  \subfloat[][]{\includegraphics[width=0.33\textwidth]{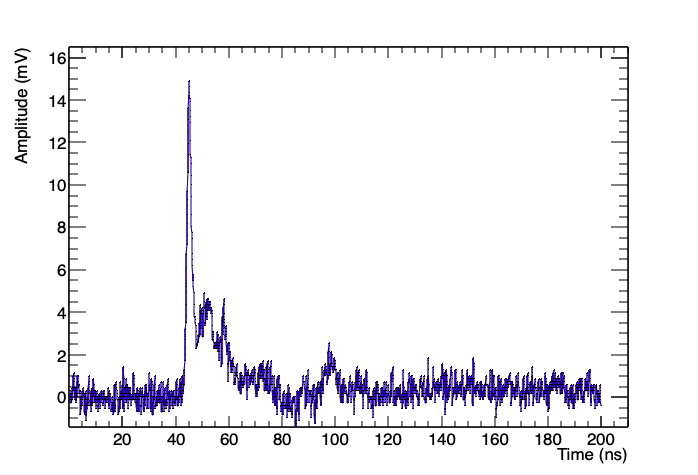} } 
   \subfloat[][]{\includegraphics[width=0.33\textwidth]{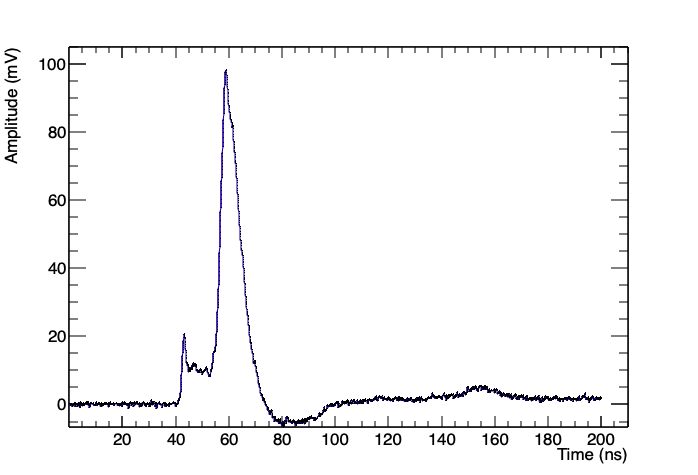} } \quad
    \caption{Signal waveforms: (a) Avalanche signal, (b) Transition charge signal and (c) Streamer signal}
    \label{fig:wave}
\end{figure}

To evaluate gas mixture performances, different types of electron prompt signals are classified as follows:

\begin{itemize}
 
 \item{Pure avalanche: it is a very short single signal, 2-3 ns FWHM in a 2 mm gap, depending on the gas mixture, which is strictly time-correlated with the trigger signal;}
 \item{Streamer: it is characterized by an avalanche signal “precursor” followed, after some very fluctuating delay, by a much larger signal lasting tens of nanoseconds. It can be interpreted as a thin column of conducting plasma interconnecting the two electrode surfaces. In the streamer event, the time interval between the avalanche precursor and the streamer appearance is characterized by a multiple avalanche regime and/or by a large tail of the avalanche precursor. The streamers occur usually at voltages above the “knee” of the efficiency plateau. In principle when operating in “avalanche mode”, the absence of streamers is desirable but this strongly depends on the gas-mixture properties;}
 \item{Transition signal: in some cases we can observe a multiple avalanche signal and/or a large tail following the precursor, which do not trigger however any streamer production. These events, therefore, are neither streamers nor avalanches.  They are classified in the following as transition signals because they are typical of a high-voltage range where the avalanche-to-streamer transition is gradually happening. These signals are negligible for the standard gas but relevant in the HFO gas mixtures.}
\end{itemize}
Representative waveforms for the described signals are shown in Figure \ref{fig:wave}.
In order to understand the physics of these different types of electrical discharge, the prompt and ionic charges are measured for each acquired waveform. They are defined as the charge integral over the whole acquired time window of 170 ns and 100 $\rm{\mu}$s for the prompt and ionic signal respectively. Moreover, the charge associated with the single avalanche signal is defined as the integrated charge over 10 ns around the avalanche peak. This time window is somewhat larger than the avalanche duration to take into account the time jitter of the trigger with respect to the signal itself. Therefore, concerning the prompt charge, we can distinguish the total prompt charge, integrated on a 170 ns window, from the single avalanche charge integrated over a 10 ns window. \\

The measured prompt-charge value and the time-over-threshold are the two parameters used for discriminating among the three kinds of events established above. 
For this purpose the threshold has been fixed to 15\% of the maximum amplitude peak in the event. 
A single avalanche is defined by a charge not exceeding 5 pC and a time-over-threshold below 12 ns, which is the typical duration of an avalanche signal in the standard mixture, increased with an adequate margin. A transition signal is defined by the logical OR of two conditions: i) a total prompt charge between 5 pC and 30 pC with a time-over-threshold exceeding 12 ns or ii) the charge integrated over the tail of the signal, excluding therefore the avalanche contribution, exceeding 21\% of the total charge associated to the event (called exceeding charge below). 
A streamer is defined by a total prompt charge above 30 pC and a time-over-threshold exceeding 30 ns.
All these conditions are summarized in Table \ref{tab:a}.

\begin{table}[ht]
  \caption{Avalanche, streamer and transition event criteria. The exceeding charge is defined as the charge integrated over the tail of the signal, excluding therefore the avalanche contribution.}
      \label{tab:a}
  \begin{tabular*}{\textwidth}{cccc}
    \hline
     Signal type  & Prompt charge  & Time over threshold & exceeding charge  \\
    \hline  
      avalanche &$ \leq$5 pC & <12 & -\\
     streamer & > 30 pC & >30 ns & -\\    

     transition event & 5 $ \leq$ q (pC)<30 &  $ \geq$ 12 ns &$ >$ 0.21 \\
    \hline
  \end{tabular*}
 \end{table}

\section{Experimental results}
\label{sec:experimental_results}

In this section the results of 4-components gas mixtures F-HFO/$\rm{CO}_{2}$/i-$\rm{C}_{4}\rm{H}_{10}$/Cl-HFO are presented. The mixtures under study have a GWP of few units.

\subsection{Test of low-GWP gas mixtures }
\label{sec:data}
The tests described in  section \ref{sec:experimental_test} suggest that a new eco-gas mixture replacing the standard one can be found according to the following criteria:
\begin{itemize}
	\item {The Tetrafluoroethane is replaced by the C$_{3}$H$_{2}$F$_{4}$/CO$_{2}$ mixture, which becomes the main component of the new gas;}
	\item{The i-C$_{4}$H$_{10}$ is kept in the eco-mixture with a concentration just below the flammability threshold. Its fraction in the following tests is  7\%. Minor readjustments depending on the exact value of the mixture flammability threshold do not  change the test results in a substantial way;}
	\item{ The SF$_{6}$ is replaced by C$_{3}$H$_{2}$F$_{3}$Cl (HFO1233zd) which has similar quenching properties with much lower GWP.}
\end{itemize}

\begin{figure}[h]
 	\centering
 	\includegraphics[width=0.9\textwidth]{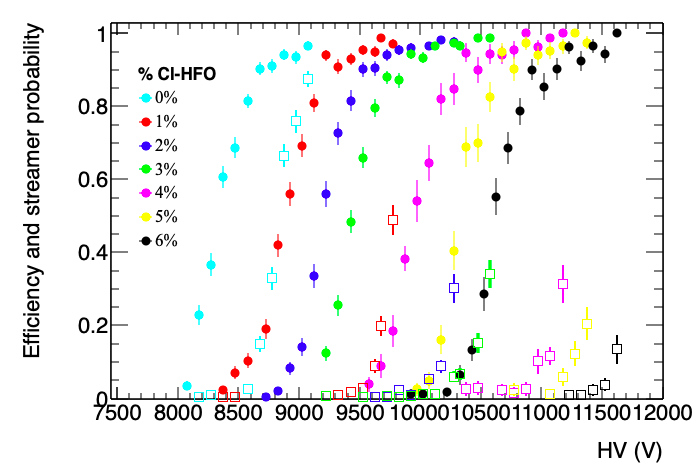}
 	\caption{Efficiency and streamer probability as a function of the high voltage for different Cl-HFO concentration. The blank markers are referred to the prompt efficiency, while the empty markers represent the streamer probability. 
}
 \label{fig:effvshvhfo1233zd}
 \end{figure}

The search for the optimal concentration of C$_{3}$H$_{2}$F$_{3}$Cl, indicated as Cl-HFO below, is the subject of this section.
We test therefore the performance of an RPC working with the 4-component gas mixture of: F-HFO/CO$_{2}$/i-C$_{4}$H$_{10}$/Cl-HFO, with -HFO/i-C$_{4}$H$_{10}$ kept at constant concentration of 15\% and 7\% respectively and CO$_{2}$/Cl-HFO ratio ranging from 77/0 to 71/6 in steps of 1\%. Detection efficiency and streamer probability as a function of the high voltage is reported in Figure \ref{fig:effvshvhfo1233zd} for all these mixtures with different CO$_{2}$/Cl-HFO ratio.
According to Figure \ref{fig:effvshvhfo1233zd} the effect of Cl-HFO on the RPC performance   can be summarized in the following points:

\begin{itemize}
	\item {The operating voltage increases at the rate of about 400 V/1\% Cl-HFO
;}
	\item{The detection efficiency at the knee of the plateau is at least 93\% for all the gas mixtures
;}
	\item{ The mixture not containing Cl-HFO shows a substantial overlapping between the operating voltages of the avalanche and streamer modes, the streamer contamination being about 35\% at the plateau knee 
 ;}
 	\item{ For all mixtures containing Cl-HFO the avalanche-to-streamer separation, defined as the voltage range between the plateau knee and the voltage at which the streamer probability is 10\%, is about 400 V.}
\end{itemize}

\begin{figure}[h]
 	\centering
 	\includegraphics[width=1.\textwidth]{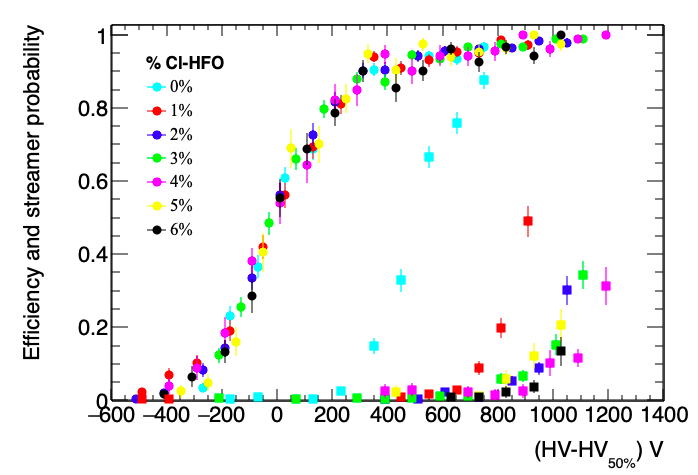}
 	\caption{Efficiency and streamer probability as a function of the high voltage for different Cl-HFO concentration. The curves have been aligned respect to the high voltage value which corresponds to the 50\% of efficiency}
 \label{fig:rialleff}
 \end{figure}

In order to evidence shape differences among different plots in Figure \ref{fig:effvshvhfo1233zd}, we aligned all plots at the high-voltage value corresponding to 50\% detection efficiency. The results reported in Figure \ref{fig:rialleff} show that all efficiency plots have the same shape. Indeed the good overlapping ($\chi^{2}$/ndf = 1.12, see Figure \ref{fig:fittone}) shows that the difference among different Cl-HFO concentrations is just a displacement along the high-voltage axis. On the contrary the streamer probability, aligned with the same rule, decreases for increasing Cl-HFO concentration. This effect however is very strong up to 2\% concentration and much weaker above, giving comparable streamer curves.

\begin{figure}[h]
 	\centering
 	\includegraphics[width=1.\textwidth]{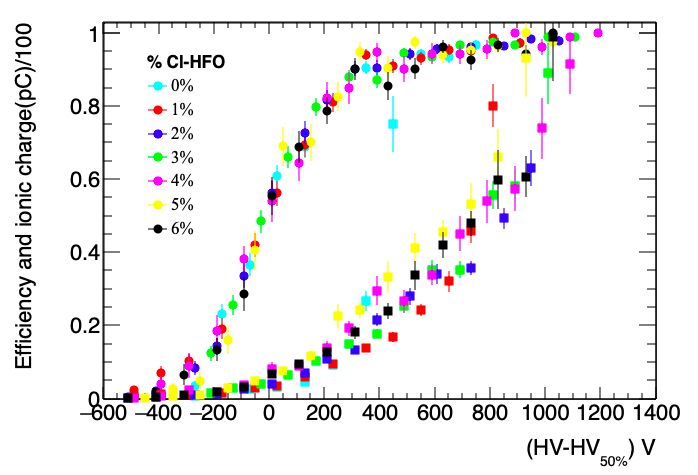}
 	\caption{Ionic charge as a function of the high voltage for different Cl-HFO concentration}
 \label{fig:data_ionicchargevshv}
 \end{figure}

In Figure \ref{fig:data_ionicchargevshv} the ionic charge as a function of the high voltage is shown, using the same shift as in the previous plots. The ionic charge of the mixture without Cl-HFO reaches very high values ($\sim$ 75 pC) at low efficiency due to premature streamer appearance, whereas the mixtures with 5\% and 6\% of Cl-HFO content have a ionic charge exceeding 30 pC at the first plateau point and reach very high values at the operating voltage, due to much higher operating field. The other mixtures show a total charge between 20 and 30 pC up to 300 V above the operating voltage and a slower rise if compared to the others. The mixture with 1\% of Cl-HFO shows the lowest ionic charge in the first three plateau points, but has a fast rise due to streamer appearance. This results suggests that the mixture with about 2\%-3\% of Cl-HFO should be the best one regarding both avalanche-streamer separation and total charge delivered inside the detector. It should be stressed that all the results reported in this section are strongly related to the charge threshold chosen for the efficiency measurements, which in this paper is 0.5 pC = 5 $\times$ $\rm{\sigma_{noise}}$, to be compared with the threshold used in the ATLAS RPCs.

\begin{figure}

\centering
  \subfloat[][]{\includegraphics[width=0.38\textwidth]{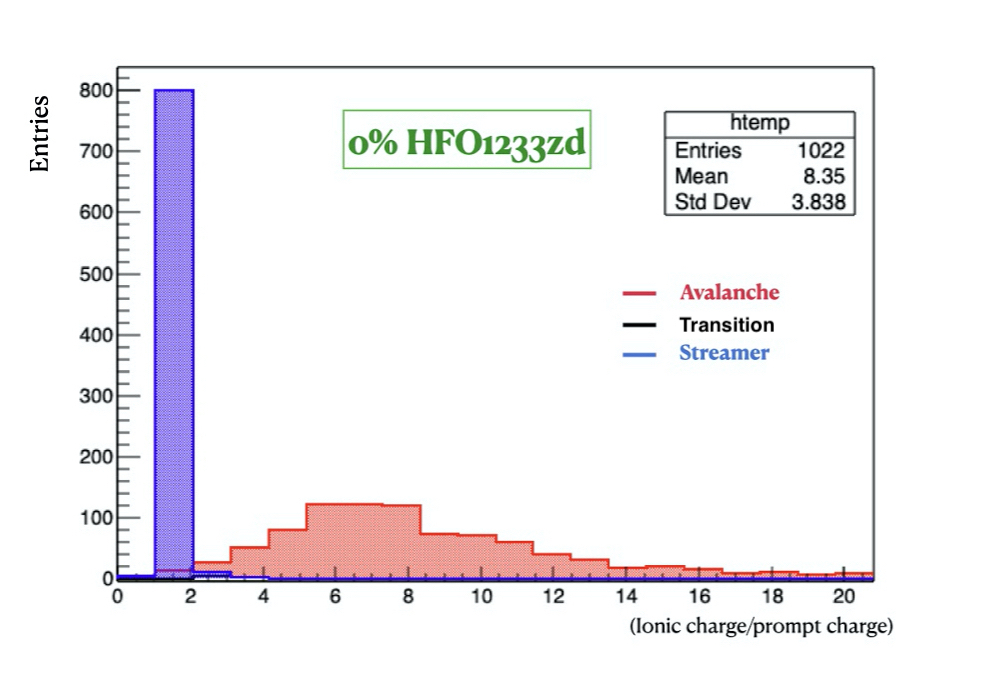} } \quad
  \subfloat[][]{\includegraphics[width=0.38\textwidth]{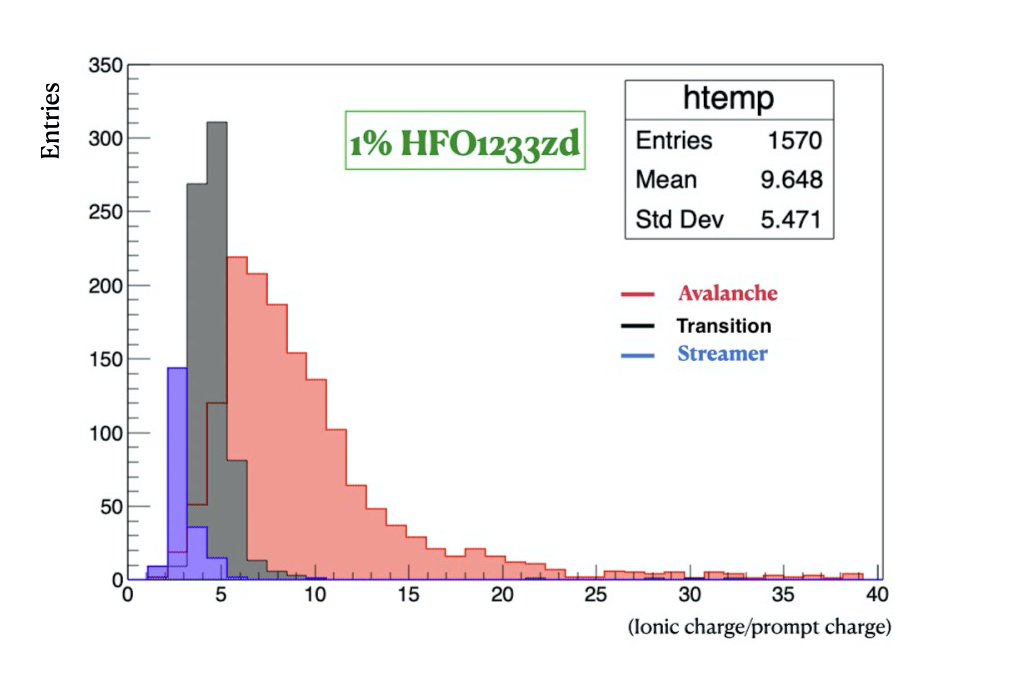} } \quad
   \subfloat[][]{\includegraphics[width=0.38\textwidth]{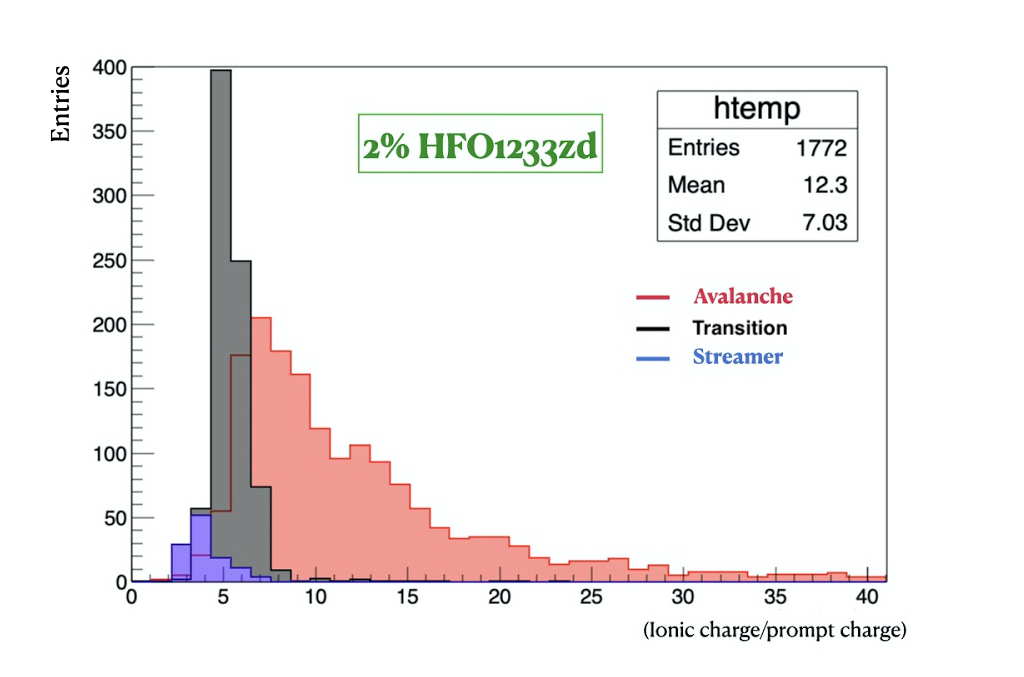} } \quad
    \subfloat[][]{\includegraphics[width=0.38\textwidth]{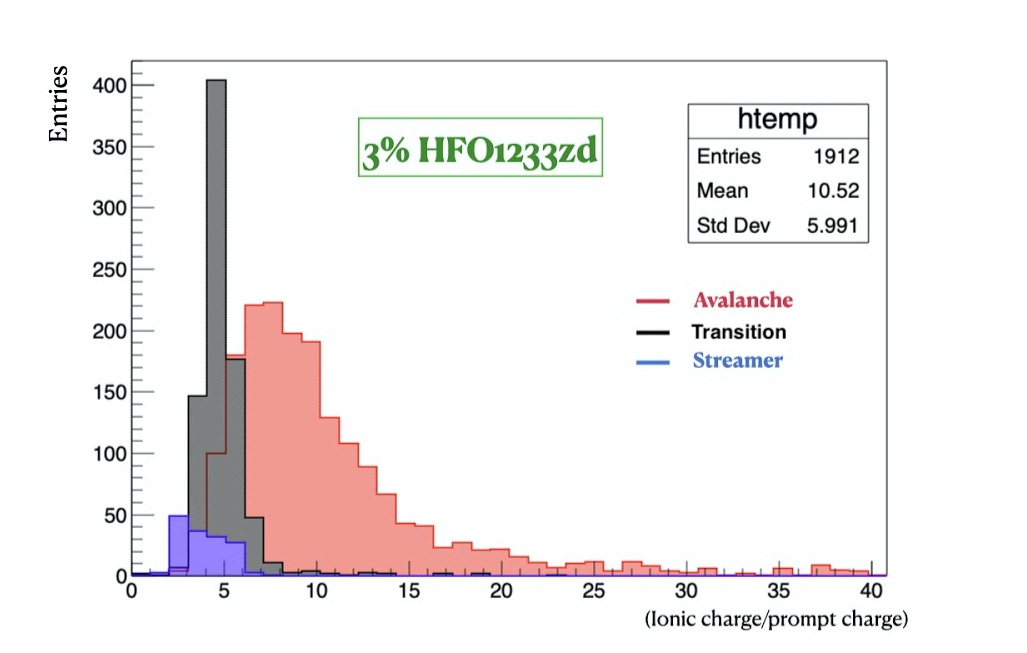} }  
     \caption{Ionic/prompt ratio of avalanches, streamers and extra charges for different Cl-HFO percentages: (a) 0\%, (b)  1\%, (c) 2\% and (d)  3\%}
    \label{fig:ratios}
\end{figure}

Figure \ref{fig:ratios} shows distributions of the ionic-to-prompt charge ratio of avalanche, streamer and transition signals for different Cl-HFO concentrations. 
These plots strongly support the classification of the observed waveforms as avalanches, streamers, and transition signals, showing that it is not arbitrary but based on different physical properties of each category. Indeed, purely exponential models of the avalanche growth give ionic/prompt =$\rm{\alpha}$g $\sim$ 10-12, $\rm{\alpha}$ and g being the Townsend coefficient and the gas gap size respectively . The interpretation of the streamer as a plasma column where electrons and ions are uniformly distributed gives ionic/prompt = 1. The addition of an electronegative quenching like Cl-HFO generates the category of the transition signals as a mixing of avalanche and streamer properties. Moreover, it also reduces, to some extent, the difference between avalanches and streamers increasing the ionic/prompt value from 1 to a few units.

\subsection{Experimental plots: Fit analysis}
\label{sec:fit}
The results described in section \ref{sec:data} show that after, aligning the high voltage scale to the  50\% efficiency point for all plots, the efficiency curves have the same profile.\\ The plot summarizing all experimental data contains  97 points and offers therefore an excellent opportunity to find an adequate fitting function. The data analysis requires indeed interpolating functions that allow correlating with continuity different quantities without the constraint of limiting the correlation to the values of the experimental points. It should be stressed that the functions we use in the fit are not based on physical models. They are just suggested by the similarity between the shape of the fitting function and the experimental plot. We use the following fitting functions:

\begin{itemize}
	\item {The detection efficiency vs HV is usually fitted with the Fermi function. In this case however the trend of the efficiency plateau shows a systematic slow rise after the knee. We used therefore a "double Fermi" function fit :

\begin{equation}
\label{eq:fit}
f(x)= \frac{p_{0}}{1+e{\frac{p_{1}-x}{p_{2}}}}+  \frac{1-p_{0}}{1+e{\frac{p_{3}-x}{p_{4}}}}
\end{equation}}

	\item{ Streamer probability vs HV : single Fermi function fit;}
 	\item{ Transition events vs HV: Landau-type fit. This fit indeed reproduces the fact that the transition-event probability grows, reaching a maximum at the first streamer appearance, and then slowly decreases as the streamer probability increases;}
  	\item{ Prompt and ionic charge vs HV: multi-degree polynomial fit.}
\end{itemize}

\begin{figure}[h]
 	\centering
 	\includegraphics[width=0.9\textwidth]{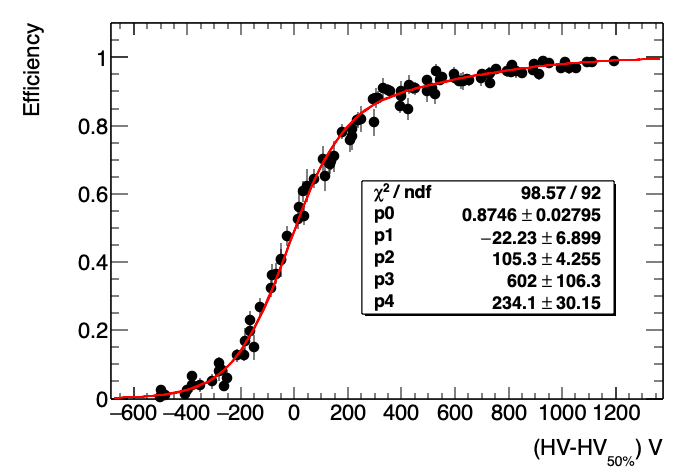}
 	\caption{Fit result on the global efficiency curve using the double-Fermi function}
 \label{fig:fittone}
 \end{figure}

The quality of the double-Fermi function fit is shown in Figure \ref{fig:fittone} where a 5-parameter function can fit 97 experimental points with an excellent $\chi^{2}$.

\subsubsection{Comparison of gas mixtures with different Cl-HFO concentrations}
\label{sec:fit_comp}
The choice among mixtures fulfilling the crucial prerequisites of non-flammability and environment-friendship is mainly based on the total charge/pulse delivered in the gas. Indeed, the lower this charge the lower the operating current and therefore the higher the rate capability. 
This is subject anyway to the condition that the signal has to pass the front-end electronics discrimination threshold in order to be detected. For the ATLAS chambers already installed, this threshold is 0.4 pC and is very close to the threshold required in this test. For future chambers the threshold will be substantially lower. 

\begin{figure}[h]
 	\centering
 	\includegraphics[width=0.8\textwidth]{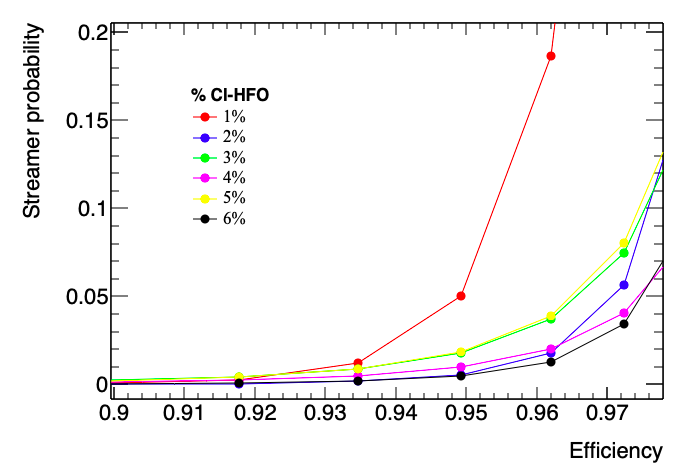}
 	\caption{Streamer probability as a function of the efficiency}
 \label{fig:streamer_fit}
 \end{figure}

 \begin{figure}[h]
 	\centering
 	\includegraphics[width=0.8\textwidth]{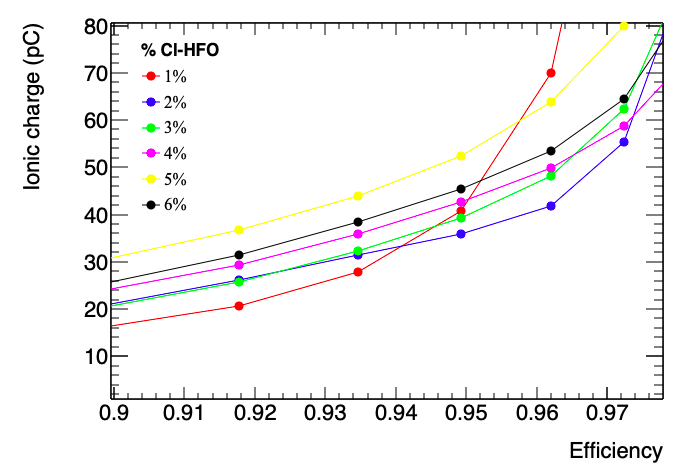}
 	\caption{Ionic charge as a function of the efficiency}
 \label{fig:ionic_fit}
 \end{figure}

The properties of gas mixtures with different Cl-HFO concentrations are compared in Table \ref{tab:percentuali} and Table \ref{tab:chargeeff}. The optimal choice of the Cl-HFO fraction is based on the criterium of minimizing both streamer fraction and total delivered charge per pulse, at efficiency $\geq$ 95\%. These data are reported in detail in Figure \ref{fig:streamer_fit} and Figure \ref{fig:ionic_fit} respectively and summarized in Table \ref{tab:percentuali} and Table \ref{tab:chargeeff} .
According to these tables the optimal concentration is around 2\%. However the range (3-6)\% gives good performance anyway taking also into account that the final choice has to include other parameters, like for example the aging, that are not yet measured.

\begin{table}[ht]
  \caption{Streamer and transition events probability for the tested mixtures at different efficiency values}
      \label{tab:percentuali}
  \begin{tabular*}{\textwidth}{c|cccccccc}
   \hline  
     & \% Cl-HFO  &0&1&2&3&4&5&6\\
    \hline 
  \multirow{2}*{Efficiency ~ 90\%}&streamer(\%)&20&0&0&0&0&0&0\\
\cline{2-2} 
&transition event (\%) &0&11&5&5&7&25&9\\
\hline
  \multirow{2}*{Efficiency ~ 95\%}&\%streamer&86&5&0&2&1&2&0\\
\cline{2-2} 
&\%transition event &0&50&45&54&57&74&60\\
\hline
  \multirow{2}*{Efficiency ~ 96\%}&\%streamer&92&20&2&4&2&4&1\\
\cline{2-2} 
&\%transition event &0&51&57&65&70&77&70\\   
    
  \end{tabular*}
 \end{table}

 \begin{table}[ht]
  \caption{Ionic and prompt charges of the tested mixtures at different efficiency values}
      \label{tab:chargeeff}
  \begin{tabular*}{\textwidth}{c|cccccccc}
   \hline  
     &\% Cl-HFO  &0&1&2&3&4&5&6\\
    \hline 
  \multirow{2}*{Efficiency ~ 90\%}&Ionic charge (pC)&45&15&20&20&23&30&24\\
&Prompt charge (pC) &27&2.4&2&2.2&2.7&4&2.5\\
\hline
  \multirow{2}*{Efficiency ~ 95\%}&Ionic charge (pC)&240&40&36&39&42&52&45\\
&Prompt charge (pC) &209&10&6&7&7&9&7\\
\hline
  \multirow{2}*{Efficiency ~ 96\%}&Ionic charge (pC)&270&70&42&48&50&62&53\\
&Prompt charge (pC)&241&19&8&9&9&11&9\\   
\hline   
  \end{tabular*}
 \end{table}

\subsection{Characterization of the new mixture and comparison with the standard one }

\begin{figure}[!h]
\centering
  \subfloat[]{\includegraphics[width=0.33\textwidth]{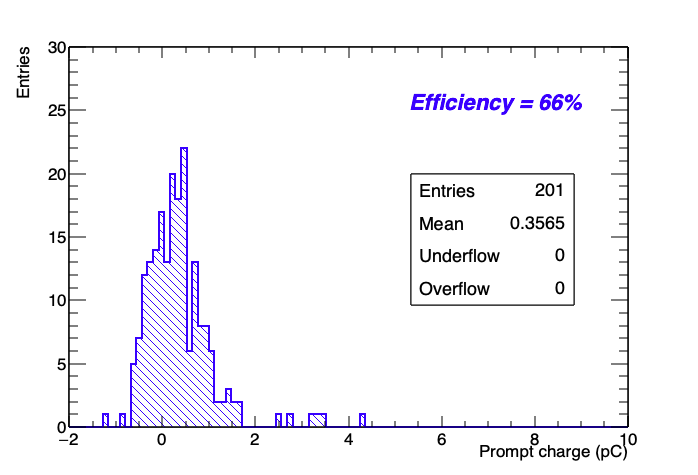} } 
  \subfloat[]{\includegraphics[width=0.33\textwidth]{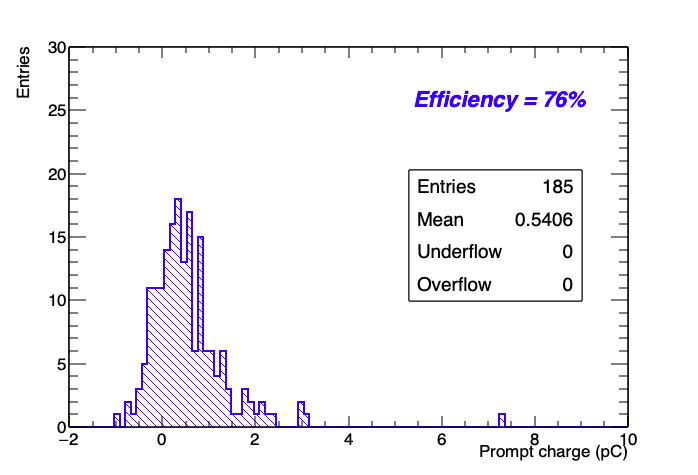} } 
  \subfloat[]{\includegraphics[width=0.33\textwidth]{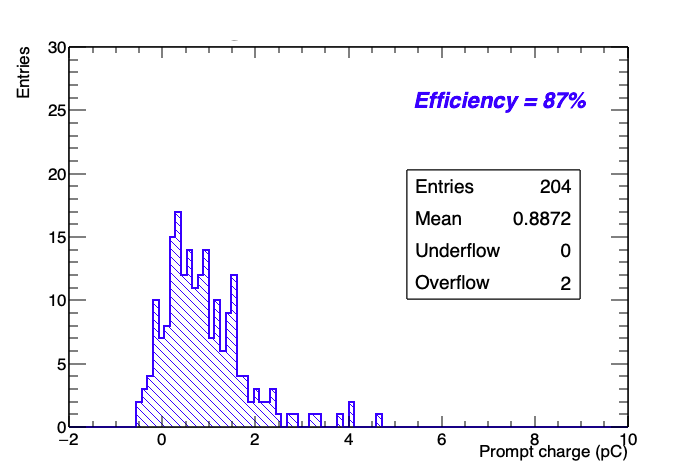} } \quad
  \subfloat[]{\includegraphics[width=0.33\textwidth]{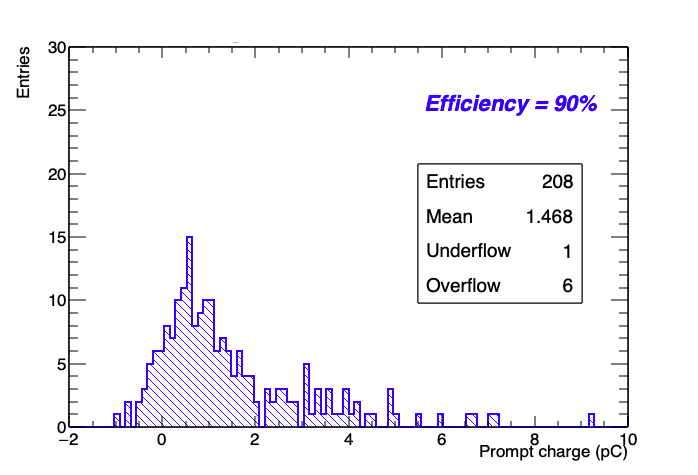} } 
  \subfloat[]{\includegraphics[width=0.33\textwidth]{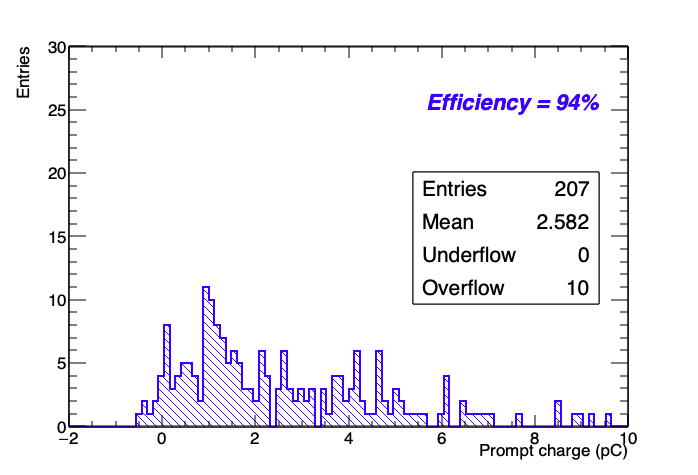} } 
  \subfloat[]{\includegraphics[width=0.33\textwidth]{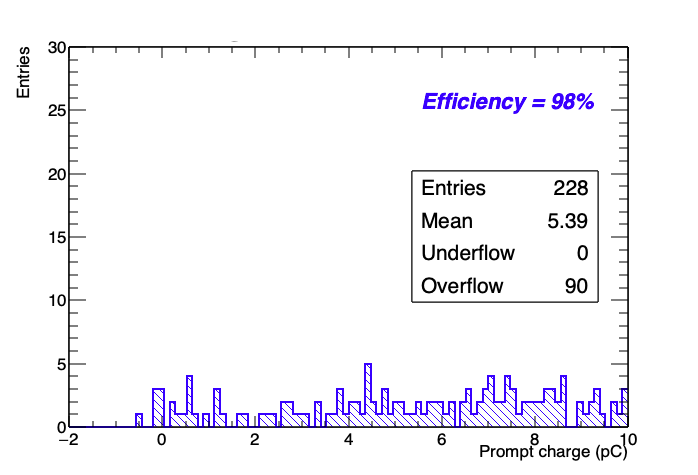} } \quad
  
    \caption{Total prompt charge distribution at different efficiencies}
    \label{fig:chadis}
\end{figure}

From the previous section the mixture with Cl-HFO = 2\% appears to be the most suitable in terms of avalanche-streamer high-voltage separation and total charge delivered inside the detector. The characteristics of this gas mixture, even if not at the same level of the standard one, allow anyway an efficient streamerless operation in avalanche mode.  The total prompt charge distribution of the eco mixture with 2\% Cl-HFO is shown in Figure \ref{fig:chadis} for different efficiency values. These distributions show a clear peak up to 90\% efficiency. For higher efficiencies the distributions become much broader due to the appearance of transition events.\\

\begin{figure}[!h]

  \subfloat[][]{\includegraphics[width=0.5\textwidth]{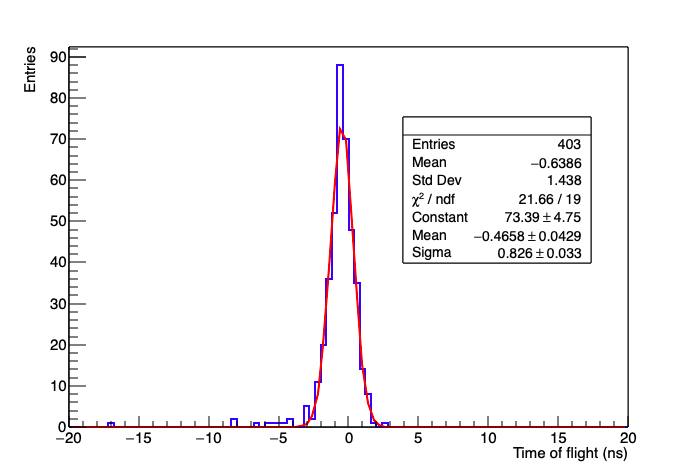} } \quad
  \subfloat[][]{\includegraphics[width=0.5\textwidth]{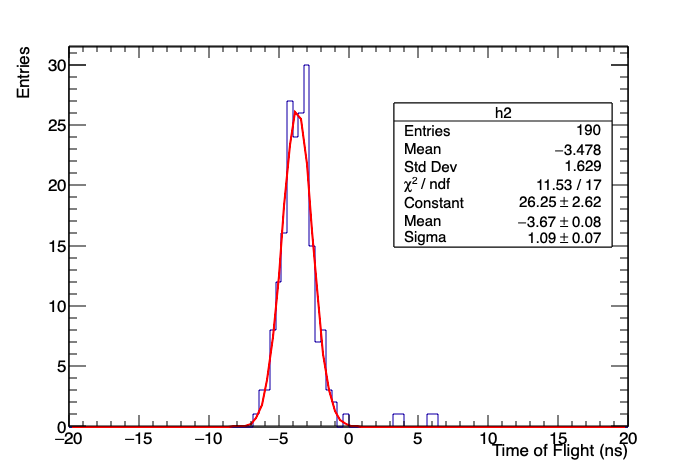} } \quad
    \caption{Time resolution for the (a) eco mixture and the (b) standard mixture }
    \label{fig:time_res}
\end{figure}

The time resolution has been measured with the time-of-flight (TOF) method, using the 0.5 mm gas gap as time reference. The time of flight is the difference between the threshold crossing time in the reference chamber and the signal peak time in the test chamber. The TOF distributions with the standard and eco mixtures are shown in Figure \ref{fig:time_res} and the corresponding time resolutions are

\begin{align*}
   \rm{ \sigma}^{Eco}_{t} = (0.83 \pm 0.03)  \rm{ns}\\
   \rm{ \sigma}^{Standard}_{t} = (1.09 \pm 0.07)  \rm{ns}
\end{align*}

These TOF distributions do not contain any kind of corrections for systematic effects due to :

\begin{itemize}
	\item {Contribution of the reference chamber time resolution;}
	\item {different propagation time of the signal in the confirm and test chamber due to the angular distribution of the cosmic rays;}
	\item{Presence of multiple peaks merged together which might broaden the signal waveform.}
\end{itemize}

The comparison between the two distributions gives anyway a time improvement of the new gas, the time resolution being the 24\% better than that of the standard mixture.

\begin{figure}[h]
 	\centering
 	\includegraphics[width=0.7\textwidth]{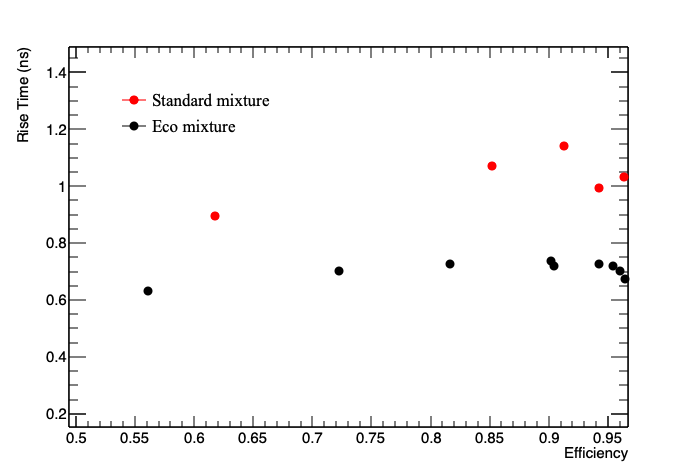}
 	\caption{Rise time as a function of the efficiency for the standard and eco mixtures. The eco mixture rise time is about 0.3 ns shorter}
 \label{fig:risetime}
 \end{figure}

The better timing properties of the new gas are confirmed by the signal rise time comparison shown in Figure \ref{fig:risetime}. The rise time is defined from 20\% to 50\% of the maximum amplitude. For fraction of the maximum amplitude > 50\%, the rise time could be affected by multi-peak events.

\section{HFO1233zd as $\pmb{\rm{SF}_{6}}$ substitute}
\label{sec:HFO_vs_SF6}
\begin{figure}[!h]

  \subfloat[][]{\includegraphics[width=0.5\textwidth]{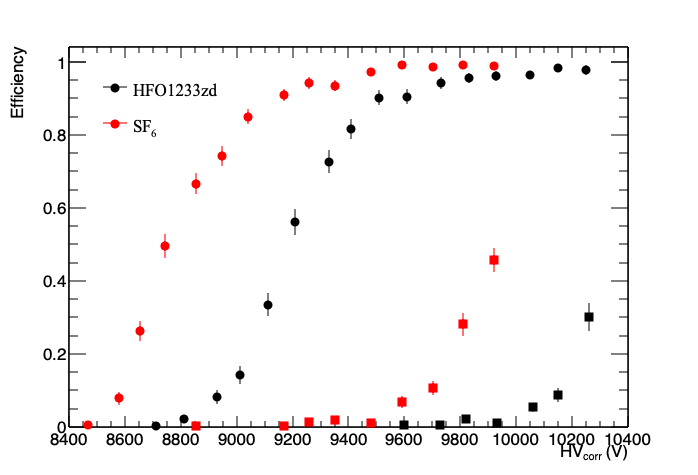} } \quad
  \subfloat[][]{\includegraphics[width=0.5\textwidth]{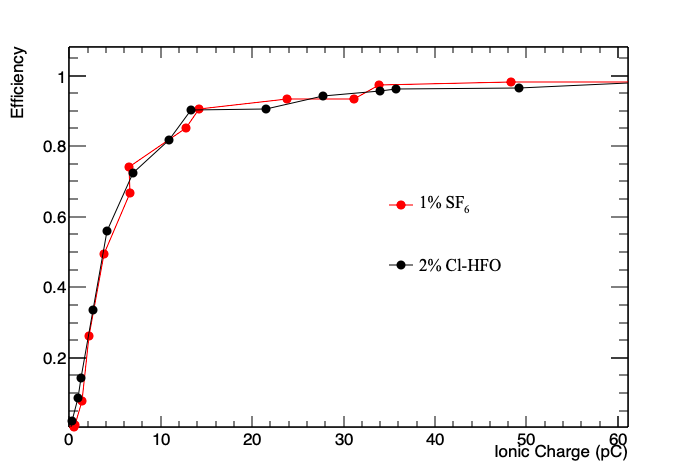} } \quad
    \caption{Efficiency as a function of the high voltage (a) Efficiency as a function of the ionic charge  (b)}
    \label{fig:hfovssf6}
\end{figure}

The experimental results suggest that Cl-HFO is a very high electronegativity gas and has properties suitable for streamer suppression. In the standard mixture and in the eco-friendly gas mixture already tested, the $\rm{SF}_{6}$ plays the role of streamer suppressor. This component has a GWP=23900, while the Cl-HFO has a GWP $\sim$7. \\
In this section we make a detailed comparison between the $\rm{SF}_{6}$ and the Cl-HFO properties. We compare mixtures composed of F-HFO/i-$\rm{C}_{4}\rm{H}_{10}$ = 15/7 and i) CO$_{2}$/ SF$_{6}$ = 77/1 or ii) CO$_{2}$/Cl-HFO = 76/2. The efficiency as a function of the high voltage is shown in Figure \ref{fig:hfovssf6}.a. The operating voltage of the mixture with Cl-HFO is 450 V higher than the one with SF$_{6}$. Both efficiency curves show a plateau knee at 90\% efficiency followed by a slow efficiency increase up to 96\% where the first streamers appear.  
The difference between the voltage knee and the streamer appearance is about $\sim$ 350 V for both mixtures and the streamer growth is steeper in the mixture with SF$_{6}$.
The efficiency curves as a function of the ionic charge (Figure \ref{fig:hfovssf6}.b) show that the two mixtures have the same ionic charge at the same efficiency value. This result suggests that the Cl-HFO can substitute the SF$_{6}$ in these gas mixtures.

\section{Conclusions}
We have tested a new eco-friendly gas mixture for RPCs: CO$_{2}$/F-HFO/i-C$_{4}$H$_{10}$/Cl-HFO=76/15/7/2 with a GWP of few units and a ODP near to zero. The test is carried out with 2 mm gap RPC. The comparison with the standard mixture  C$_{2}$H$_{2}$F$_{4}$/i-C$_{4}$H$_{10}$/SF$_{6}$= 94.7/5/0.3 (GWP $\sim$1400) shows that:

\begin{itemize}
	\item {The operating voltage interval separating the avalanche from the streamer working mode, although somewhat shorter for the new mixture, is anyway sufficient to insure a streamerless operation in avalanche mode;}
	\item{The composition of the new mixture is not critical with respect to its components. In particular 1\% concentration changes for the i-butane (depending on the exact flammability limit not yet measured) and an increase of the F-HFO concentration from 15\% to 25\% (depending on the ageing properties not yet measured) would not produce drastic changes in the mixture properties;}
	\item{ A crucial component of the standard mixture, the SF$_{6}$ molecule characterized by one of the highest GWP = 23900, can be replaced by the Cl-HFO molecule with a GWP near to unity. This new component has a boiling point of 19 °C, very close to the room temperature, and its use requires therefore some precautions. However its low concentration, around 2\%, does not produce any constraint for the mixture temperature, that can be lowered down to -10 °C without any risk of liquefaction;}
 	\item{ The new gas shows interesting timing properties, its time resolution being about 30\% better than the one of the standard gas.}
\end{itemize}

We have tested the new gas with a 2 mm gap RPC. This work should be completed by studying the new gas properties vs the gas gap size. In particular thinner gaps might require to increase the F-HFO fraction to compensate for the thinner gaseous target. Moreover an accurate study of the ageing properties is needed, which might also suggest some readjustment of the HFO-component concentration. An exact determination of the flammability limit is also required.

\end{document}